\documentclass{article}

\usepackage[T1]{fontenc}
\usepackage[utf8]{inputenc}
\usepackage{ismir}
\usepackage{amsmath,cite,url}
\usepackage{graphicx}
\usepackage{color}
\usepackage{booktabs}
\usepackage{multirow}
\usepackage{subcaption}

\usepackage{xcolor}
\usepackage{ftnxtra}
\usepackage{wrapfig}

\title{Harmonic Representations in Symbolic Music Transformers are Localized and Controllable}

\threeauthors
  {Adi Silberschein} {Weizmann Institute of Science 
  \\  \small\texttt{adi.silberschein@weizmann.ac.il}
}
  {Megan Wei} {Brown University 
  \\ \small\texttt{meganwei@brown.edu}
  }
  {Tamar Rott Shaham} {MIT CSAIL 
  \\ \small\texttt{tamarott@mit.edu}
  }

\def\authorname{A. Silberschein, M. Wei, and T. Rott Shaham}

\begin{document}

\maketitle

\begin{abstract}
Transformer models trained on symbolic music have become remarkably capable. However, it is yet unclear how they represent music structure, and whether understanding their internal representations can be used to unlock methods for model control. We examine this through a case study of harmonic analysis: when a music transformer navigates a cadence, what does it internally ``know'' about its harmonic context, and can that knowledge be used to affect the generation? Using Aria, a LLaMA-based symbolic music transformer evaluated on Bach chorales, we find that harmonic decisions are causally localized progressively from middle layers onward. Linear probes show that mode becomes decodable in the same layer range where patching identifies growing causal effect, while key identity emerges later.
We then confirm these representations are actionable for music generation control: 
steering via latent directions corresponding to a specific harmonic change reliably produces a deceptive or minor resolution, while preserving structural coherence and generation quality. This demonstrates that, at inference time, the model's internal harmonic knowledge can be directly manipulated to control the harmonic output.

\end{abstract}

\section{Introduction}

Generative models trained on symbolic music have achieved impressive performance in generating coherent and stylistically consistent compositions. Recent systems are capable of producing multi-voice harmonizations~\cite{deepbach,coconet,musictransformer, museformer, nime2025_84,wu2023chordconditionedmelodyharmonizationcontrollable,zhou2023choirtransformergeneratingpolyphonic,Benetatos2020BachDuetAD} and long-range structure~\cite{wang2024wholesong, zhouzheng2025personalizable, qu2025mupt, midigpt, wu2023compembellish}. Yet whether their success arises from structured, interpretable representations of musical concepts or from distributed and opaque ones remains unclear. In this work, we focus on harmony as a concrete lens through which to examine these questions: when a music transformer navigates a cadence, what does it internally know about its harmonic context, and where in the network is this information represented?

Understanding internal representations can also benefit model control. In many applications, it is desirable to steer generation toward specific musical properties, enforcing harmonic constraints, modulating between keys, or altering stylistic elements. However, existing control methods often rely on fine-tuning~\cite{wang2025notagenadvancingmusicalitysymbolic, midillm} or operate at the input or decoding level~\cite{deepbach,thickstun2023anticipatory}. If models possess internal knowledge of harmonic concepts, this knowledge could in principle be identified and directly manipulated at inference time, enabling targeted control without retraining.

Here, we provide evidence that the internal representations of a symbolic music transformer encode harmonic context in a structured and localized form, and that intervening on these representations enables control over harmonic decisions during generation. We study Aria~\cite{aria}, a LLaMA-based symbolic music transformer, evaluated on Bach chorales~\cite{bachchorales,boulangerlewandowski2012modelingtemporaldependencieshighdimensional}. To analyze and manipulate its internal computations, we leverage tools from interpretability, including causal interventions that localize where harmonic decisions are computed and targeted manipulations that alter these representations at generation.

Bach chorales provide an ideal testbed for this analysis: they are harmonically rich, governed by well-understood conventions, and admit precise counterfactual construction. By modifying a single note at a cadence, shifting the resolution toward a different harmonic target while keeping the surrounding context fixed, we obtain contrastive pairs that differ in exactly one musically meaningful dimension. This controlled setting enables precise causal analysis of harmonic computation within the model.

Our analysis yields three main findings. First, causal interventions reveal that harmonic decisions emerge progressively from middle layers onward, suggesting that harmonic computation is not present from the start but develops through the network. Second, linear probes trained on the residual stream show that mode (major vs. minor) becomes decodable in middle layers, while key identity emerges later, a dissociation that suggests a processing hierarchy from coarse tonal category to specific tonal center. Third, we demonstrate that these representations are actionable: by amplifying latent directions corresponding to harmonic changes, we can reliably alter the model's generated resolutions, inducing deceptive (V $\rightarrow$ vi) or minor-mode (V $\rightarrow$ i) cadences while preserving overall musical coherence. We further show that the geometry of these internal representations reflects known structures of tonal space in music theory. Audio samples, code, and data are available on our webpage\footnote{\url{https://adisilber.github.io/harmonic-interp/}}.

These findings indicate that a symbolic music transformer can encode tonal harmony in a form that is both interpretable and manipulable, suggesting that the internal representations of music generation models can serve as controls for shaping musical structure at inference time.

\section{Related Work}

\paragraph*{Music Generation Models.}
Music generation models have made rapid progress, producing high-fidelity audio across genres and styles~\cite{jukebox, musicgen, musiclm, evans2024stableaudioopen, gong2025acestep, audioldm2-2024taslp}. On the symbolic side, early neural approaches modeled Bach chorales \cite{deepbach, coconet}. Music Transformer~\cite{musictransformer} laid the ground for transformer-based symbolic generation, inspiring work on tokenization schemes that better capture musical structure~\cite{popmusictransformer, hsiao2021compound}, controllability via conditioning~\cite{figaro, musecoco2023}, and new architectures for long-range structure~\cite{museformer}. Recent work has pushed symbolic generation further, scaling to larger models, extending to richer musical contexts, and adding new sources of control~\cite{thickstun2023anticipatory, bhandari2025text2midi, wang2025notagenadvancingmusicalitysymbolic, aria, midillm, qu2025mupt, yuan2024chatmusician}. Understanding the inner mechanisms of these models could unlock more intuitive methods of inference-time control for musicians and researchers.

\paragraph*{Interpretability in Music Models.}
Linear probing \cite{alain2018understanding, belinkov2022probing} has emerged as a tool for understanding internal representations of pretrained models, and has been applied to music models. Castellon et al.~\cite{Castellon2021-ry} show that Jukebox representations transfer to MIR tasks such as genre and key, and self-supervised models such as MERT~\cite{mertLi2023-dg} and MusicFM~\cite{Won2023-sa} are likewise evaluated by linear probing of frozen representations. Wei et al.~\cite{Wei2024-music} introduce SynTheory and probe Jukebox and MusicGen for core music-theoretic concepts, finding they are encoded in deeper layers, while Koo et al.~\cite{Koo_undated-hp} probe MusicGen attention heads across tasks, including key detection, finding that some musical concepts are localized to a small subset of heads. 
Recent work has taken steps toward causal intervention and steering: \cite{velezinntermechanismsmusic} uses interchange interventions to study how MusicGen composes instrument and genre concepts; \cite{ma2024exploring} adds probe-derived vectors to MusicGen's residual stream to intervene on chord representations; \cite{smitn} steers in a feedback loop to control instrument presence; \cite{facchiano2025activationpatchinginterpretablesteering} constructs steering vectors from hidden states to control tempo and timbre; \cite{singh2025discovering} applies sparse autoencoders to MusicGen to discover and steer interpretable musical concepts; \cite{zhaosteering2026} derives steering directions via Recursive Feature Machines, controlling music theory concepts; \cite{jiang2025composer} steers composer style in symbolic music models; and TADA \cite{staniszewski2026tada} extends activation patching and steering to audio diffusion models, controlling instruments, genre, and mood. 

While prior work has probed for static chord and key attributes~\cite{Wei2024-music, ma2024exploring}, none has studied how the model handles dynamic harmonic decisions at structural moments like cadences. Our use of naturalistic contrastive pairs grounded in music theory (minimal edits of real chorales), rather than synthetic stimuli \cite{Wei2024-music}, enables precise causal analysis of harmonic decisions.

\section{Experimental Setup}

\paragraph*{Model.}
We study \emph{Aria}~\cite{aria}, a LLaMA-based transformer trained on symbolic music sequences. The model operates autoregressively over tokenized musical inputs and is evaluated without additional fine-tuning.

\paragraph*{Dataset and Counterfactual Construction.}
We evaluate the model on Bach chorales~\cite{bachchorales,boulangerlewandowski2012modelingtemporaldependencieshighdimensional}, which provide a structured and harmonically rich testbed. Chorales follow well-defined tonal conventions and contain clear cadential resolutions, making them particularly suitable for controlled analysis of harmonic behavior. We focus on harmonic structure at cadences, short sequences that establish and resolve harmonic tension. 
Cadences are extracted using harmonic analysis of music21~\cite{music21}.
To isolate harmonic decisions, we construct minimal contrastive pairs consisting of an original sequence \( x \) and a counterfactual variant \( x' \). Given a chorale containing a cadence, we modify a single note at the point of resolution to shift the harmonic outcome (e.g., toward a different tonal resolution), while keeping the surrounding context fixed. This yields pairs of sequences that differ in exactly one musically meaningful dimension. 
This controlled setup enables precise causal analysis: by comparing model behavior across original and counterfactual sequences, and by applying interventions between them, we can localize and manipulate the representations responsible for harmonic decisions. In all experiments, we focus on the model’s prediction at the cadence resolution token. 

\section{Methods}

\paragraph*{Causal Interventions.}
Interchange intervention tests causal relationships between a model’s internal representations and its behavior~\cite{vig2020investigating, geiger2021causal, meng2022locating, wang2023interpretability}. Let \( f(\cdot) \) denote the model, and let \( \mathbf{h}_\ell(x) \) denote the internal representation at layer \( \ell \) produced during a forward pass on input \( x \).
Given an \textit{original} input \( x \) and a corresponding \textit{counterfactual} input \( x' \), an interchange intervention replaces \( \mathbf{h}_\ell(x) \) with \( \mathbf{h}_\ell(x') \) while keeping the remainder of the computation unchanged, yielding an intervened output:
\begin{equation}
\hat{y}_{\text{int}} = f\big(x;\, \mathbf{h}_\ell \leftarrow \mathbf{h}_\ell(x')\big),
\end{equation}
the operation underlying what is commonly called \emph{activation patching}~\cite{zhang2024towards}.
If the intervened output \( \hat{y}_{\text{int}} \) matches the outcome associated with the counterfactual input, this provides evidence that the intervened representation \( \mathbf{h}_\ell \) plays a causal role in the model’s computation. We quantify this effect using interchange intervention accuracy (IIA)~\cite{geiger2022inducing}. In our case, since harmony continuations are determined by several generated notes, we use the IIA to score each patched continuation by its first generated note: a success if its pitch class belongs only to the counterfactual resolution chord, a failure if only to the original one, and not counted otherwise. IIA is the fraction of successes among counted samples. See Sec.~\ref{sec:couterfact} for the counterfactual construction.

\paragraph*{Linear Probing.}
To assess what harmonic information is represented across layers, we train linear probes~\cite{alain2018understanding,belinkov2022probing,hewitt-manning-2019-structural} on hidden states to predict attributes such as mode and key:
\begin{equation}
\hat{y} = \arg\max_c \; w_c^\top h^{(\ell)} + b_c,
\end{equation}
where \( h^{(\ell)} \) is a model internal representation vector and \( c \) indexes harmonic attributes. Probe accuracy reflects how well the attribute is linearly decodable from the representation, allowing us to track where harmonic information becomes accessible in the network. We use the trained probe weights to identify linear directions in the representation space that correspond to specific harmonic attributes.

\paragraph*{Representation Steering.}
To test whether representations are actionable, we perform interventions along a direction 
$v \in \mathbf{R}^d$ in representation space ~\cite{turner2024steeringlanguagemodelsactivation, rimsky-etal-2024-steering}:
\begin{equation}
\tilde{h}_t^{(\ell)} = h_t^{(\ell)} + \alpha v,
\end{equation}
where \( \alpha \) controls the strength of the intervention. We evaluate how this modification affects the generated sequence to test whether \( v \) corresponds to a meaningful harmonic transformation. Our steering direction represents the subspaces encoding harmonic information.

\begin{figure}[t]
    \centering
    \includegraphics[width=1\linewidth]{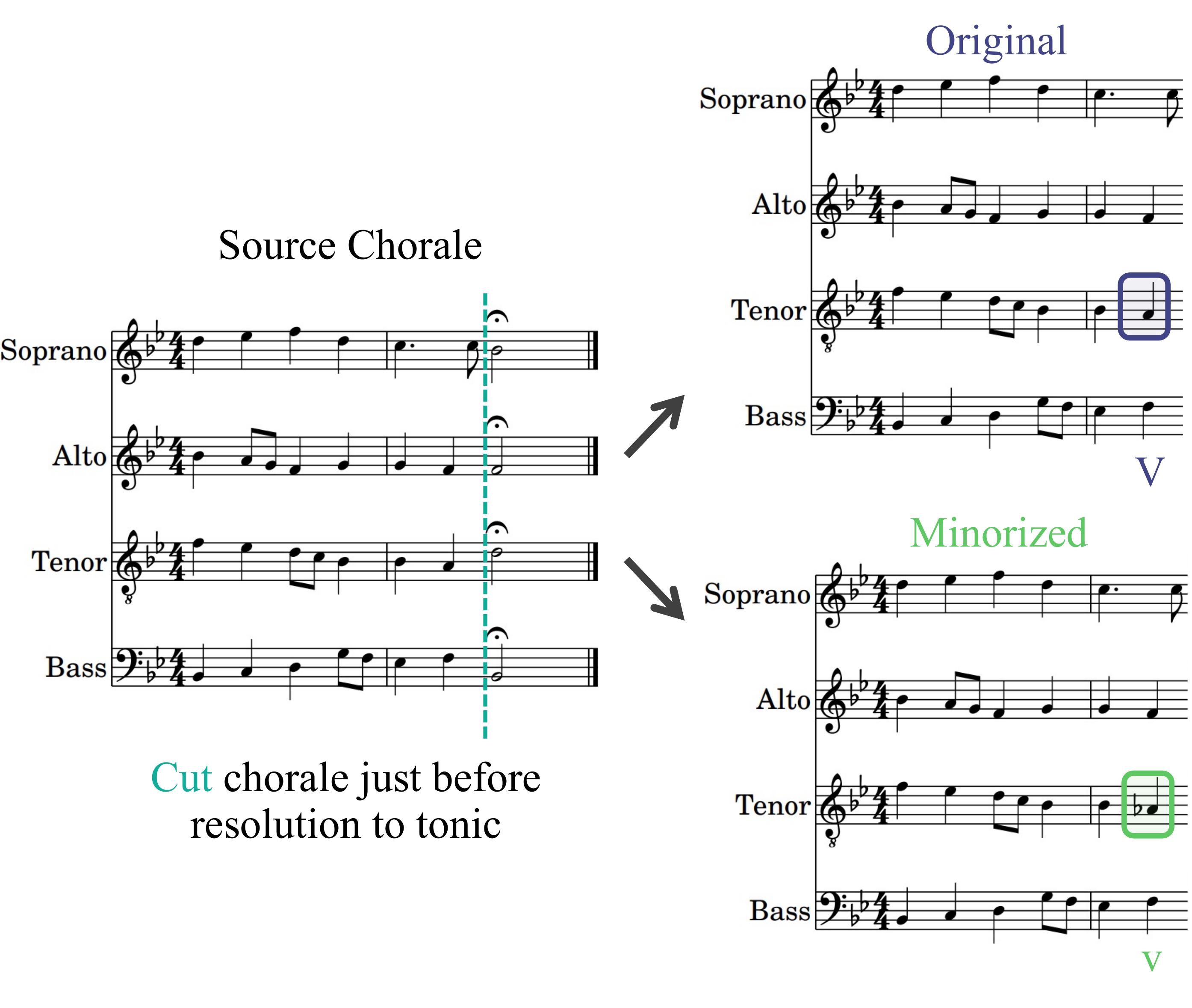}
    \caption{\textbf{Local Intervention.} A V$\rightarrow$I cadence is identified in a Bach chorale (left). The chorale is truncated at the end of the V chord, and the chordal third of V is lowered by a semitone, turning V into v (right), yielding an original-counterfactual pair that differs in exactly one note. }
\label{fig:local_intervention}
\end{figure}

\section{Experiments}

We evaluate our approach through three complementary analyses: causal interventions to localize harmonic computation, linear probing to characterize how harmonic information is encoded, and activation steering to demonstrate direct control over harmonic output at inference time.

\subsection{Where Does the Model Compute Harmonic Representations?}

\paragraph*{Local Intervention on Bach Chorales.} \label{sec:couterfact}
To study how Aria processes harmonic information, we construct a set of contrastive pairs from Bach chorales~\cite{bachchorales,boulangerlewandowski2012modelingtemporaldependencieshighdimensional}. 
Our goal is to isolate a single harmonic change while keeping all other context identical, enabling precise causal analysis.
We identify cadences with V$\rightarrow$I resolutions in major chorales and V$\rightarrow$i in minor chorales using music21~\cite{music21}. We then truncate each chorale at the end of the V chord and lower a single pitch (the chordal third of V) by a semitone, turning the major dominant (V) into a minor dominant (v), as shown in Fig.~\ref{fig:local_intervention}; we refer to the edited version as \textit{minorized}. This manipulation is minimal (single pitch change), yet it redirects the harmonic expectation. We apply this intervention to both major and minor source chorales, yielding $184$ and $91$ original/counterfactual prompt pairs, each continued with three sampling seeds ($552$ and $273$ generation pairs).

\begin{figure}[t]
    \centering
    \includegraphics[width=1\linewidth]{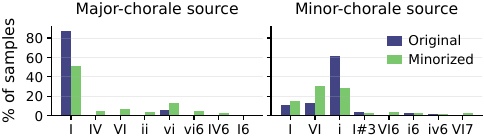}
    \caption{\textbf{Intervention on a single pitch shifts cadence resolution.} Modifying a single pitch in a V$\rightarrow$I or V$\rightarrow$i cadence substantially redistributes the model's resolution distribution in both major and minor source chorales.}

    \label{fig:input_intervention}
\end{figure}

\paragraph*{A Single Pitch Shifts Cadence Resolution.}
We first ask whether this minimal intervention is sufficient to affect the model's harmonic behavior. We feed both the original and the intervened versions to Aria and record how it resolves the cadence. As shown in Fig.~\ref{fig:input_intervention}, the intervention substantially shifts the resolution distribution. In major-source chorales, resolutions to the tonic (I) drop from $87.5\%$ to $51.4\%$, with mass redistributing toward the submediant (vi) and other chords. In minor-source chorales, the original resolves predominantly to the minor tonic (i; ${\sim}60\%$), while the minorized version substantially reduces this, redistributing mass toward other chords. These results confirm that a single pitch change at a cadence is sufficient to redirect the model's harmonic output.

\begin{figure}[t]
        \centering    \includegraphics[height=0.45\linewidth, width=1\linewidth]{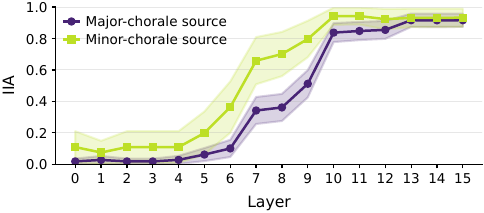}
    \caption{\textbf{Causal Effect of Harmonic Information.} Patching residual stream representations from minorized into the original sequences at each layer causes an increase in IIA beginning in the middle layers, indicating that harmonic computation develops progressively through the network. Shaded regions denote 95\% confidence intervals.}

    \label{fig:activation_patching}
\end{figure}

\paragraph*{Causal Effect of Harmonic Information Emerges in Middle Layers.}
We next perform activation patching (interchange intervention) using each original-intervened pair. For each layer separately, we patch the residual stream activation of the last token before generation from the intervened sequence into the original, and measure how much the model's output shifts toward the intervened resolution. As shown in Fig.~\ref{fig:activation_patching}, early layers (0--5) carry little causal harmonic information, as patching them has minimal effect. An increased causal effect emerges around layer 6, and continues to grow through later layers. This suggests that harmonic computation is not present from the start but develops progressively, with middle layers marking the presence of causally relevant harmonic encoding.

\subsection{How is Harmonic Information Encoded?}

We investigate whether the harmonic representations from the residual stream are linearly decodable. We test two hypotheses: the model encodes the \textit{mode} of the chorale (major vs.\ minor, $2$ classes) and the \textit{key} of the chorale ($2$ modes $\times 12$ notes = $24$ classes, of which only $18$ occur in the corpus). For each, we train a linear probe. We use the original train-test split from~\cite{bachchorales,boulangerlewandowski2012modelingtemporaldependencieshighdimensional}, and extract activations from the residual stream of the last token of each bar in the training set, yielding  $3,176$ samples. We train a separate probe for each layer.

\begin{figure}[t]
    \centering
    \includegraphics[
    height=0.45\linewidth, 
    width=1\linewidth]{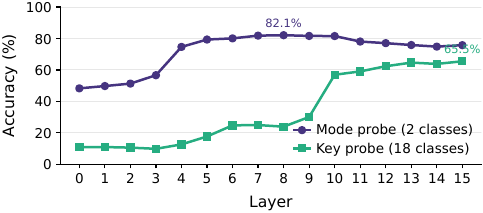}
    \caption{\textbf{Linear probe accuracy across layers. } Mode (chance = 50\%) becomes decodable in the middle layers (from layer 4), while key identity (chance = 5.6\%) emerges later in the network (from layer 10). Note that classes might be imbalanced.}

        \label{fig:probe_acc}
\end{figure}

\paragraph*{Mode is Decodable in Middle Layers.}
As shown in Fig.~\ref{fig:probe_acc}, mode classification accuracy rises sharply from chance (50\%) in early layers, reaching a peak of 82.1\% around layer 8--9, before declining slightly in later layers. This suggests that middle layers are most specialized for encoding tonal category; the model's representation of whether it is operating in a major or minor context is strongest at this point in the network.

\paragraph*{Key Identity Emerges Later.}
Key identity (18 classes, chance $= 5.6\%$) follows a different trajectory. Probe accuracy remains low through the middle layers, then rises sharply around layer 10, reaching 65.5\% by layer 15. This late and steep emergence suggests a processing hierarchy: the model first establishes tonal \textit{category} (mode) before resolving the specific tonal \textit{center} (key).

\subsection{Can Resolution be Directly Controlled through Internal Harmonic Representations?}

Having identified that harmonic information is encoded in a decodable form in middle layers, we now ask whether these representations can be manipulated to steer the model's harmonic output at inference time. In this section, we focus on steering major chorales as a V$\rightarrow$I major cadence serves as a natural case for harmonic modifications.

\paragraph*{Steering Method.}
We derive steering directions from the trained probes. Let $w_{c_1}$ and $w_{c_2}$ denote the probe weight vectors trained to classify two target classes (e.g. two keys or major vs.\ minor mode) from the residual stream of layer $\ell$. We define the normalized steering vector at a layer $\ell$ as:
\begin{equation}
    v_\ell = \frac{\overline{\|h_\ell\|} \cdot (w_{c_1} - w_{c_2})}{\|w_{c_1} - w_{c_2}\|}
\end{equation}
where $\overline{\|h_\ell\|}$ is the mean hidden state norm of layer $\ell$ computed once from the corpus, scaling $v_\ell$ relative to the typical magnitude of the hidden states. We then apply the intervention by adding this vector, scaled by a factor $\alpha$, to the residual stream activations. We apply this intervention at all layers $\ell$ where probe accuracy is highest, corresponding to the distinct peaks observed in Fig.~\ref{fig:probe_acc} (layers 11--15 for key steering and 5--10 for mode steering). For each steering condition, we scan a range of $\alpha$ values and report the one yielding the strongest harmonic effect. We find that steering all token positions, including generated ones, yields the strongest and most consistent shifts, though all positional settings produce qualitatively similar effects, suggesting that harmonic representations are active globally throughout the sequence rather than concentrated at a single position.

\paragraph*{Evaluating Musical Qualities.}
We evaluate the quality of steered generations across four dimensions, reported in Table~\ref{tab:steering_eval}. (a) \textit{Sequence-level} metrics capture structural integrity: structural error rate (malformed \texttt{<pitch, onset, duration>} token triples), degenerate pitch repetition (runs of $\geq$4 identical pitches), and average generated sequence duration. (b) \textit{Note-level} metrics capture local statistics: out-of-range notes ($\geq$3 notes outside the normal chorales range), average note duration, and pitch variance across the generation. (c) \textit{Perceptual Quality} is measured using Fréchet Music Distance (FMD) \cite{retkowski2024frechet} computed against Bach's original Chorales. (d) \textit{Harmony} is assessed via two metrics: Roman numeral (RN) entropy, the Shannon entropy of the chord distribution over the first 16 sixteenth-note slots of the generated continuation (higher = more harmonically varied), and top-1 dominance, the fraction of the first bar occupied by the single most common Roman numeral (lower = less degenerate repetition). Together, these metrics allow us to verify that steering shifts harmonic behavior without degrading structural integrity or musical quality.

\begin{figure}[t]
    \centering
    \includegraphics[height=0.5\linewidth, width=1\linewidth]{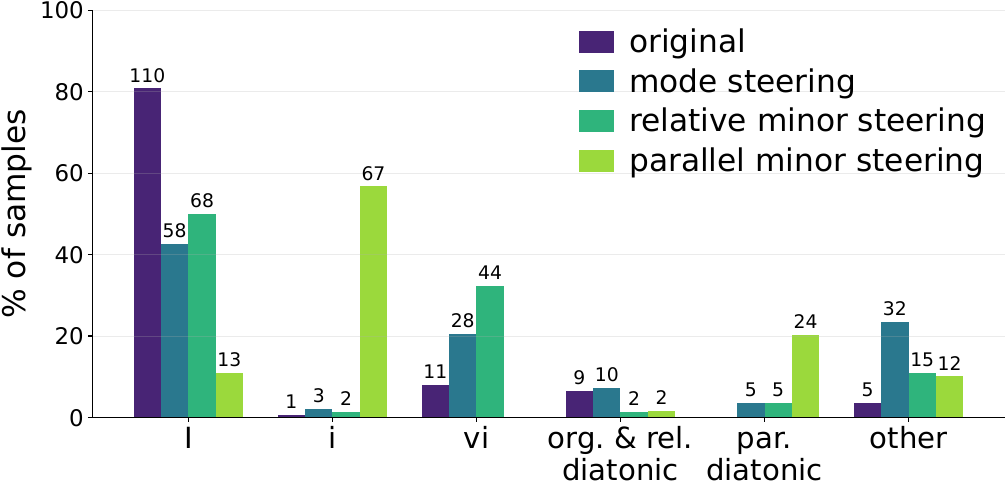}
    \caption{\textbf{Key and Mode Steering.} Applying steering on major chorales with steering directions corresponding to relative and parallel minor keys substantially increases the frequency of alternative resolutions compared to the unsteered baseline, which usually resolves to the major tonic (I). Relative minor steering shifts mass toward the submediant (vi), as it is the tonic of the relative minor, while parallel minor steering produces the most decisive shift toward the parallel minor tonic (i). Mode steering also reduces I resolutions but produces no structured redistribution, without concentrating on a specific harmonic target. Numbers above bars are counts per category. \emph{org.\ \& rel.\ diatonic}: the chords diatonic to the original key and its relative minor; \emph{par.\ diatonic}: chords of the parallel minor only; \emph{other}: all remaining resolutions.
    }

\label{fig:key_steering}
\end{figure}

\paragraph*{Mode Steering Produces Diffuse Redistribution.}
As shown in Fig.~\ref{fig:key_steering}, mode steering (optimal with $\alpha=0.25$) does reduce resolutions to the tonic (I; from 110 to 58 cases), confirming that the mode representation carries some causal influence over harmonic output. However, unlike key steering, the redirected mass does not concentrate on a specific harmonic target, but instead spreads broadly across alternative categories. In addition, while key steering benefits from intervening across all token positions, including generated tokens, mode steering achieves its best results when applied only at the last bar positions, suggesting that mode information plays a limited role in shaping harmonic decisions at generation.

\paragraph*{Key Steering Shifts Cadence Resolution.}
As shown in Fig.~\ref{fig:key_steering}, key steering substantially redirects the model's harmonic output. Without steering, resolutions are overwhelmingly to the tonic (I; $81\%$). Applying a relative minor direction (with $\alpha=0.15$) shifts mass strongly toward the submediant (vi; $32\%$), as it is the tonic of the relative minor. Applying a parallel minor direction ($\alpha=0.1$) produces a different but equally interpretable outcome, shifting mass toward the parallel minor tonic (i; $57\%$ cases). Both interventions produce musically meaningful harmonic changes. As confirmed by Table~\ref{tab:steering_eval}, both interventions preserve generation quality: structural errors remain below 3\%, FMD stays relatively close to the unsteered baseline, and RN entropy increases slightly, indicating that steered generations are more harmonically varied rather than collapsing onto a single chord. Together, these results demonstrate that harmonic behavior can be directly controlled at inference time without retraining and without degrading musical quality.

\begin{table*}[h]
\centering
\small
\begin{tabular}{lccccccccc}
\toprule
& \multicolumn{3}{c}{Sequence-Level}
& \multicolumn{3}{c}{Note-Level} &
Quality & \multicolumn{2}{c}{Harmony} \\
\cmidrule(lr){2-4} \cmidrule(lr){5-7} \cmidrule(lr){8-8} \cmidrule(lr){9-10}
& Struct. & Pitch & Avg.\ gen.\ & Out of & Note dur. & Pitch & FMD & RN & Top-1 \\
& error & repeat & duration & range & avg & var. & & entropy & dom.\ \\
& (\%)$\downarrow$ & (\%)$\downarrow$ & (ms) & (\%)$\downarrow$ & (ms) & & $\downarrow$ & (bits)$\uparrow$ & (\%)$\downarrow$ \\
\midrule
Aria (no steering)
    & 0.0 & 0.0 & 3740
    & 0.0 & 735 & 60
    & 80.27 & 1.26 & 59.0 \\
Relative minor steering
    & 2.9 & 0.0 & 3800
    & 0.0 & 728 & 55
    & 85.42 & 1.40 & 55.3 \\
Parallel minor steering
    & 0.0 & 0.0 & 3783
    & 0.0 & 741 & 55
    & 106.65 & 1.50 & 49.6 \\
\bottomrule
\end{tabular}
\caption{\textbf{Key Steering Evaluation.} Results confirm that generation quality is well-preserved across key steering conditions: structural errors remain below 3\% and out-of-range notes near 0\%, FMD stays close to the unsteered baseline (FMD calculated against original Bach chorales; for reference, the FMD between Bach chorales and Bach fugues is 195.97, far larger than any steering-induced change), and RN entropy varies by at most 0.24\,bits while top-1 dominance remains near 50\% --- showing the steered model generates harmonically varied continuations rather than collapsing onto a single chord. For $\downarrow$ lower is better, for $\uparrow$ higher is better, when no indication similarity to the baseline is preferred.}
\label{tab:steering_eval}
\end{table*}

\begin{figure}[t]
    \centering
    \includegraphics[height=0.5\linewidth, width=1.0\linewidth]{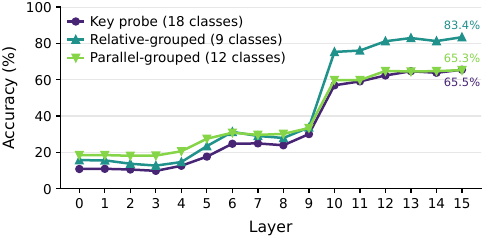}
    \caption{\textbf{Relative vs. Parallel probing.} Grouping keys by their shared key signature (relative-grouped) yields substantially higher peak accuracy (83.4\%) than the baseline 18-class key probe (65.5\%) or grouping by parallel keys (65.3\%). The probe thus separates key classes more easily when each class is defined by a shared pitch collection (relative grouping) than by a shared root (parallel grouping).}
    \label{fig:group_prob}
\end{figure}

\begin{figure*}[t]
    \centering
    \includegraphics[width=1\textwidth]{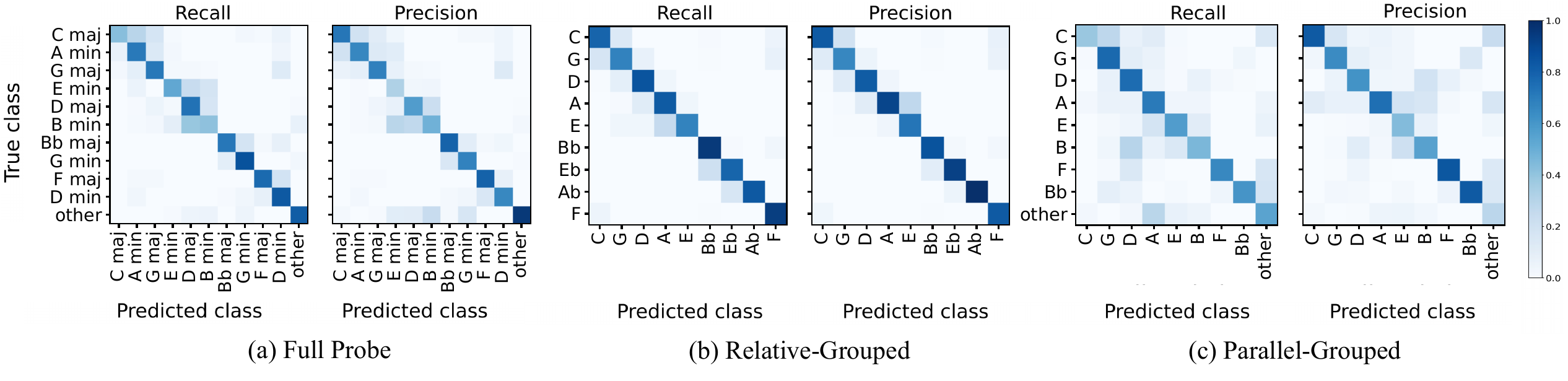}
     \caption{\textbf{Relative keys are more confusable than parallel keys.}
     Confusion matrices at probe peak accuracy layer. (a) The full probe shows systematic confusions between relative key pairs (e.g., F major with D minor, Bb major with G minor), indicating that the model might place relative keys close together in representation space. (b) Grouping by shared key signature (relative-grouped) resolves these confusions, yielding a near-diagonal matrix. (c)~Grouping by shared tonal center (parallel-grouped) retains more off-diagonal mass, confirming that parallel keys are more distinct. 
    }
    \label{fig:confussion}
\end{figure*}

\subsection{Why is Parallel Steering More Effective than Relative Steering?}
The steering experiments revealed an asymmetry: parallel minor steering ($\alpha=0.1$) produced cleaner and more decisive harmonic shifts than relative minor steering ($\alpha=0.15$), despite requiring a smaller intervention. Surprisingly, parallel minor steering can be achieved by intervening on a single layer (layer 12) while getting equivalent steering effect and generation quality ($76$ out of $118$ test cases shift to harmonic resolution toward target key with FMD=$102.05$), whereas relative minor steering requires the full layer range to produce comparable shifts. This raises a natural question: why does the choice of target key matter, and what does it reveal about how the model internally organizes tonal relationships?

\paragraph*{Probing Tonal Relationships.}
To investigate how the model organizes tonal space internally, we train two additional linear probes. The \textit{relative-grouped} probe collapses the 18 keys into 9 classes by shared key signature (e.g., C major and A minor share the same class), while the \textit{parallel-grouped} probe collapses them into 12 classes by shared tonal center (e.g., C major and C minor). As shown in Fig.~\ref{fig:group_prob}, the relative-grouped probe achieves higher peak accuracy (83.4\%) than both the full probe (65.5\%) and the parallel-grouped probe (65.3\%). This suggests that grouping the key classes according to key signature made the classes more separable, indicating that the model places relative keys closer together in representation space than parallel keys, making them less distinguishable and harder to steer between.
\paragraph*{The Model Confuses Relative but not Parallel Keys.}
The confusion matrices shown in Fig.~\ref{fig:confussion} extend this pattern. In the full 18-class probe (Fig.~\ref{fig:confussion}(a)), the dominant errors occur between relative pairs (e.g., C major is confused with A minor) while parallel keys are distinguished more cleanly. The relative-grouped confusion matrix (Fig.~\ref{fig:confussion}(b)) is nearly perfectly diagonal: once relative pairs are collapsed into the same class, the remaining confusions largely disappear, confirming that relative key internal representation similarity accounts for most of the full probe's errors. The parallel-grouped matrix (Fig.~\ref{fig:confussion}(c)) retains more off-diagonal mass even after collapsing parallel pairs, indicating a different source of similarity. 
This explains the steering asymmetry: steering between relative keys means pushing in a direction the model already finds ambiguous, producing less decisive shifts. Steering between parallel keys moves along a direction that is clean and well-separated in the model's representational space, making a smaller intervention ($\alpha=0.1$) sufficient to produce dramatic harmonic shifts.

\paragraph*{Internal Geometry Mirrors Music Theory.}
The internal geometry of Aria aligns with the structure of tonal space in music theory. Relative keys share the same key signature and diatonic pitch collection, differing only in their tonal center. Their musical boundary is inherently ambiguous: composers routinely exploit this ambiguity through pivot chord modulations. Parallel keys share the same tonal center but differ in their pitch collection, key signature, and harmonic vocabulary. The shift between them is unambiguous and immediately perceptible. 

\section{Conclusion}
We investigated how tonal harmony is represented and controlled in Aria, a LLaMA-based symbolic music transformer. Harmonic computation emerges progressively from middle layers onward, where mode becomes decodable before key identity. Steering via latent directions derived from key probes redirects cadence resolution while preserving musical coherence, demonstrating harmonic control at inference time. The geometry of these representations mirrors music theory: relative keys lie closer together than parallel keys in representation space. Our analysis is grounded entirely in Bach chorales, which are more harmonically constrained than broader symbolic music corpora. Whether it holds for less conventional or more chromatic material remains open for future work.

\newpage
\section{Acknowledgement}
TRS is supported by ARL, MIT-IBM Watson AI Lab, and Hyundai Motor Company. 

\section{AI Usage Statement}

Large language models were used to assist in polishing portions of this paper's text, and AI coding agents were used to assist in the development of experimental code. All content, including the scientific claims, experimental results, and references, was created and is the responsibility of the authors.

\section{Ethics Statement}
This work investigates mechanistic interpretability and controllability of a symbolic music transformer, with the goal of better understanding how such models encode and use musical knowledge. The primary contribution is analytical, we do not release a new generative system or deploy one in a public-facing context.
The ability to manipulate a model's internal representations to steer its outputs may provide fine-grained generative control that could be valuable for composers, music technologists, and accessibility tools. 

\bibliography{main}

\begin{thebibliography}{10}
\providecommand{\url}[1]{#1}
\csname url@samestyle\endcsname
\providecommand{\newblock}{\relax}
\providecommand{\bibinfo}[2]{#2}
\providecommand{\BIBentrySTDinterwordspacing}{\spaceskip=0pt\relax}
\providecommand{\BIBentryALTinterwordstretchfactor}{4}
\providecommand{\BIBentryALTinterwordspacing}{\spaceskip=\fontdimen2\font plus
\BIBentryALTinterwordstretchfactor\fontdimen3\font minus \fontdimen4\font\relax}
\providecommand{\BIBforeignlanguage}[2]{{%
\expandafter\ifx\csname l@#1\endcsname\relax
\typeout{** WARNING: IEEEtran.bst: No hyphenation pattern has been}%
\typeout{** loaded for the language `#1'. Using the pattern for}%
\typeout{** the default language instead.}%
\else
\language=\csname l@#1\endcsname
\fi
#2}}
\providecommand{\BIBdecl}{\relax}
\BIBdecl

\bibitem{deepbach}
G.~Hadjeres, F.~Pachet, and F.~Nielsen, ``{DeepBach}: A steerable model for {B}ach chorales generation,'' in \emph{Proc. of the 34th Int. Conf. on Machine Learning (ICML)}, vol.~70, Sydney, Australia, 2017, pp. 1362--1371.

\bibitem{coconet}
C.-Z.~A. Huang, T.~Cooijmans, A.~Roberts, A.~Courville, and D.~Eck, ``Counterpoint by convolution,'' in \emph{Proc. of the 18th Int. Society for Music Information Retrieval Conf. (ISMIR)}, Suzhou, China, 2017, pp. 211--218.

\bibitem{musictransformer}
C.-Z.~A. Huang, A.~Vaswani, J.~Uszkoreit, I.~Simon, C.~Hawthorne, N.~Shazeer, A.~M. Dai, M.~D. Hoffman, M.~Dinculescu, and D.~Eck, ``{Music Transformer}: Generating music with long-term structure,'' in \emph{Proc. of the Int. Conf. on Learning Representations (ICLR)}, New Orleans, LA, USA, 2019.

\bibitem{museformer}
B.~Yu, P.~Lu, R.~Wang, W.~Hu, X.~Tan, W.~Ye, S.~Zhang, T.~Qin, and T.-Y. Liu, ``{Museformer}: Transformer with fine- and coarse-grained attention for music generation,'' in \emph{Advances in Neural Information Processing Systems (NeurIPS)}, vol.~35, 2022, pp. 1376--1388.

\bibitem{nime2025_84}
L.~Blanchard, C.~Holt, and J.~A. Paradiso, ``{AI} harmonizer: Expanding vocal expression with a generative neurosymbolic music {AI} system,'' in \emph{Proc. of the Int. Conf. on New Interfaces for Musical Expression (NIME)}, Canberra, Australia, 2025, pp. 578--581.

\bibitem{wu2023chordconditionedmelodyharmonizationcontrollable}
S.~Wu, X.~Li, and M.~Sun, ``Chord-conditioned melody harmonization with controllable harmonicity,'' in \emph{Proc. IEEE Int. Conf. on Acoustics, Speech and Signal Processing (ICASSP)}, Rhodes Island, Greece, 2023, pp. 1--5.

\bibitem{zhou2023choirtransformergeneratingpolyphonic}
J.~Zhou, H.~Zhu, and X.~Wang, ``{Choir Transformer}: Generating polyphonic music with relative attention on transformer,'' \emph{arXiv preprint arXiv:2308.02531}, 2023.

\bibitem{Benetatos2020BachDuetAD}
C.~Benetatos, J.~VanderStel, and Z.~Duan, ``{BachDuet}: A deep learning system for human-machine counterpoint improvisation,'' in \emph{Proc. of the Int. Conf. on New Interfaces for Musical Expression (NIME)}, Birmingham, UK, 2020, pp. 635--640.

\bibitem{wang2024wholesong}
Z.~Wang, L.~Min, and G.~Xia, ``Whole-song hierarchical generation of symbolic music using cascaded diffusion models,'' in \emph{Proc. of the Int. Conf. on Learning Representations (ICLR)}, Vienna, Austria, 2024.

\bibitem{zhouzheng2025personalizable}
C.~Zhou-Zheng and P.~Pasquier, ``Adaptable symbolic music infilling with {MIDI-RWKV},'' \emph{arXiv preprint arXiv:2506.13001}, 2025.

\bibitem{qu2025mupt}
X.~Qu, Y.~Bai, Y.~Ma, Z.~Zhou, K.~M. Lo, J.~Liu, R.~Yuan, L.~Min, X.~Liu, T.~Zhang, X.~Du, S.~Guo, Y.~Liang, Y.~Li, S.~Wu, J.~Zhou, T.~Zheng, Z.~Ma, F.~Han, W.~Xue, G.~Xia, E.~Benetos, X.~Yue, C.~Lin, X.~Tan, W.~Huang, J.~Fu, and G.~Zhang, ``{MuPT}: A generative symbolic music pretrained transformer,'' in \emph{Proc. of the Int. Conf. on Learning Representations (ICLR)}, Singapore, 2025.

\bibitem{midigpt}
P.~Pasquier, J.~Ens, N.~Fradet, P.~Triana, D.~Rizzotti, J.-B. Rolland, and M.~Safi, ``{MIDI-GPT}: A controllable generative model for computer-assisted multitrack music composition,'' in \emph{Proc. of the AAAI Conf. on Artificial Intelligence (AAAI)}, vol.~39, no.~2, Philadelphia, PA, USA, 2025, pp. 1474--1482.

\bibitem{wu2023compembellish}
S.-L. Wu and Y.-H. Yang, ``{Compose \& Embellish}: Well-structured piano performance generation via a two-stage approach,'' in \emph{Proc. IEEE Int. Conf. on Acoustics, Speech and Signal Processing (ICASSP)}, Rhodes Island, Greece, 2023, pp. 1--5.

\bibitem{wang2025notagenadvancingmusicalitysymbolic}
Y.~Wang, S.~Wu, J.~Hu, X.~Du, Y.~Peng, Y.~Huang, S.~Fan, X.~Li, F.~Yu, and M.~Sun, ``{NotaGen}: Advancing musicality in symbolic music generation with large language model training paradigms,'' in \emph{Proc. of the 34th Int. Joint Conf. on Artificial Intelligence (IJCAI)}, Montreal, Canada, 2025, pp. 10\,207--10\,215.

\bibitem{midillm}
S.-L. Wu, Y.~Kim, and C.-Z.~A. Huang, ``{MIDI-LLM}: Adapting large language models for text-to-{MIDI} music generation,'' in \emph{NeurIPS 2025 Workshop on AI for Music: Where Creativity Meets Computation}, San Diego, CA, USA, 2025.

\bibitem{thickstun2023anticipatory}
J.~Thickstun, D.~Hall, C.~Donahue, and P.~Liang, ``{Anticipatory Music Transformer},'' \emph{Transactions on Machine Learning Research}, 2024.

\bibitem{aria}
L.~Bradshaw, A.~Spangher, H.~Fan, S.~Biderman, and S.~Colton, ``Scaling self-supervised representation learning for symbolic piano performance,'' in \emph{Proc. of the 26th Int. Society for Music Information Retrieval Conf. (ISMIR)}, Daejeon, South Korea, 2025, pp. 451--459.

\bibitem{bachchorales}
M.~Allan and C.~K.~I. Williams, ``Harmonising chorales by probabilistic inference,'' in \emph{Advances in Neural Information Processing Systems (NeurIPS)}, vol.~17, 2004, pp. 25--32.

\bibitem{boulangerlewandowski2012modelingtemporaldependencieshighdimensional}
N.~Boulanger-Lewandowski, Y.~Bengio, and P.~Vincent, ``Modeling temporal dependencies in high-dimensional sequences: Application to polyphonic music generation and transcription,'' in \emph{Proc. of the 29th Int. Conf. on Machine Learning (ICML)}, Edinburgh, Scotland, UK, 2012.

\bibitem{jukebox}
P.~Dhariwal, H.~Jun, C.~Payne, J.~W. Kim, A.~Radford, and I.~Sutskever, ``{Jukebox}: A generative model for music,'' \emph{arXiv preprint arXiv:2005.00341}, 2020.

\bibitem{musicgen}
J.~Copet, F.~Kreuk, I.~Gat, T.~Remez, D.~Kant, G.~Synnaeve, Y.~Adi, and A.~D{\'e}fossez, ``Simple and controllable music generation,'' in \emph{Advances in Neural Information Processing Systems (NeurIPS)}, vol.~36, 2023, pp. 47\,704--47\,720.

\bibitem{musiclm}
A.~Agostinelli, T.~I. Denk, Z.~Borsos, J.~Engel, M.~Verzetti, A.~Caillon, Q.~Huang, A.~Jansen, A.~Roberts, M.~Tagliasacchi, M.~Sharifi, N.~Zeghidour, and C.~Frank, ``{MusicLM}: Generating music from text,'' \emph{arXiv preprint arXiv:2301.11325}, 2023.

\bibitem{evans2024stableaudioopen}
Z.~Evans, J.~D. Parker, C.~Carr, Z.~Zukowski, J.~Taylor, and J.~Pons, ``{Stable Audio Open},'' in \emph{Proc. IEEE Int. Conf. on Acoustics, Speech and Signal Processing (ICASSP)}, Hyderabad, India, 2025, pp. 1--5.

\bibitem{gong2025acestep}
J.~Gong, Y.~Song, W.~Zhao, S.~Wang, S.~Xu, J.~Guo, and X.~Yang, ``{ACE-Step} 1.5: Pushing the boundaries of open-source music generation,'' \emph{arXiv preprint arXiv:2602.00744}, 2026.

\bibitem{audioldm2-2024taslp}
H.~Liu, Y.~Yuan, X.~Liu, X.~Mei, Q.~Kong, Q.~Tian, Y.~Wang, W.~Wang, Y.~Wang, and M.~D. Plumbley, ``{AudioLDM} 2: Learning holistic audio generation with self-supervised pretraining,'' \emph{IEEE/ACM Transactions on Audio, Speech, and Language Processing}, vol.~32, pp. 2871--2883, 2024.

\bibitem{popmusictransformer}
Y.-S. Huang and Y.-H. Yang, ``{Pop Music Transformer}: Beat-based modeling and generation of expressive pop piano compositions,'' in \emph{Proc. of the 28th ACM Int. Conf. on Multimedia (MM)}, Seattle, WA, USA, 2020, pp. 1180--1188.

\bibitem{hsiao2021compound}
W.-Y. Hsiao, J.-Y. Liu, Y.-C. Yeh, and Y.-H. Yang, ``{Compound Word Transformer}: Learning to compose full-song music over dynamic directed hypergraphs,'' in \emph{Proc. of the AAAI Conf. on Artificial Intelligence (AAAI)}, vol.~35, no.~1, Virtual, 2021, pp. 178--186.

\bibitem{figaro}
D.~{von R{\"u}tte}, L.~Biggio, Y.~Kilcher, and T.~Hofmann, ``{FIGARO}: Controllable music generation using learned and expert features,'' in \emph{Proc. of the Int. Conf. on Learning Representations (ICLR)}, Kigali, Rwanda, 2023.

\bibitem{musecoco2023}
P.~Lu, X.~Xu, C.~Kang, B.~Yu, C.~Xing, X.~Tan, and J.~Bian, ``{MuseCoco}: Generating symbolic music from text,'' \emph{arXiv preprint arXiv:2306.00110}, 2023.

\bibitem{bhandari2025text2midi}
K.~Bhandari, A.~Roy, K.~Wang, G.~Puri, S.~Colton, and D.~Herremans, ``{Text2midi}: Generating symbolic music from captions,'' in \emph{Proc. of the AAAI Conf. on Artificial Intelligence (AAAI)}, vol.~39, no.~22, Philadelphia, PA, USA, 2025, pp. 23\,478--23\,486.

\bibitem{yuan2024chatmusician}
R.~Yuan, H.~Lin, Y.~Wang, Z.~Tian, S.~Wu, T.~Shen, G.~Zhang, Y.~Wu, C.~Liu, Z.~Zhou, L.~Xue, Z.~Ma, Q.~Liu, T.~Zheng, Y.~Li, Y.~Ma, Y.~Liang, X.~Chi, R.~Liu, Z.~Wang, C.~Lin, Q.~Liu, T.~Jiang, W.~Huang, W.~Chen, J.~Fu, E.~Benetos, G.~Xia, R.~Dannenberg, W.~Xue, S.~Kang, and Y.~Guo, ``{ChatMusician}: Understanding and generating music intrinsically with {LLM},'' in \emph{Findings of the Association for Computational Linguistics: ACL 2024}, Bangkok, Thailand, 2024, pp. 6252--6271.

\bibitem{alain2018understanding}
G.~Alain and Y.~Bengio, ``Understanding intermediate layers using linear classifier probes,'' in \emph{Proc. of the Int. Conf. on Learning Representations (ICLR), Workshop Track}, Toulon, France, 2017.

\bibitem{belinkov2022probing}
Y.~Belinkov, ``Probing classifiers: Promises, shortcomings, and advances,'' \emph{Computational Linguistics}, vol.~48, no.~1, pp. 207--219, 2022.

\bibitem{Castellon2021-ry}
R.~Castellon, C.~Donahue, and P.~Liang, ``Codified audio language modeling learns useful representations for music information retrieval,'' in \emph{Proc. of the 22nd Int. Society for Music Information Retrieval Conf. (ISMIR)}, Online, 2021, pp. 88--96.

\bibitem{mertLi2023-dg}
Y.~Li, R.~Yuan, G.~Zhang, Y.~Ma, X.~Chen, H.~Yin, C.~Xiao, C.~Lin, A.~Ragni, E.~Benetos, N.~Gyenge, R.~Dannenberg, R.~Liu, W.~Chen, G.~Xia, Y.~Shi, W.~Huang, Z.~Wang, Y.~Guo, and J.~Fu, ``{MERT}: Acoustic music understanding model with large-scale self-supervised training,'' in \emph{Proc. of the Int. Conf. on Learning Representations (ICLR)}, Vienna, Austria, 2024.

\bibitem{Won2023-sa}
M.~Won, Y.-N. Hung, and D.~Le, ``A foundation model for music informatics,'' in \emph{Proc. IEEE Int. Conf. on Acoustics, Speech and Signal Processing (ICASSP)}, Seoul, Korea, 2024, pp. 1226--1230.

\bibitem{Wei2024-music}
M.~Wei, M.~Freeman, C.~Donahue, and C.~Sun, ``Do music generation models encode music theory?'' in \emph{Proc. of the 25th Int. Society for Music Information Retrieval Conf. (ISMIR)}, San Francisco, CA, USA, 2024, pp. 680--687.

\bibitem{Koo_undated-hp}
J.~Koo, G.~Wichern, F.~G. Germain, S.~Khurana, and J.~{Le Roux}, ``Understanding and controlling generative music transformers by probing individual attention heads,'' in \emph{IEEE ICASSP 2024 Workshop on Explainable Machine Learning for Speech and Audio (XAI-SA)}, Seoul, Korea, 2024.

\bibitem{velezinntermechanismsmusic}
M.~A. {V{\'e}lez V{\'a}squez}, C.~Pouw, J.~A. Burgoyne, and W.~H. Zuidema, ``Exploring the inner mechanisms of large generative music models,'' in \emph{Proc. of the 25th Int. Society for Music Information Retrieval Conf. (ISMIR)}, San Francisco, CA, USA, 2024, pp. 791--798.

\bibitem{ma2024exploring}
W.~Ma and G.~Xia, ``Exploring the internal mechanisms of music {LLM}s: A study of root and quality via probing and intervention techniques,'' in \emph{ICML 2024 Workshop on Mechanistic Interpretability}, Vienna, Austria, 2024.

\bibitem{smitn}
J.~Koo, G.~Wichern, F.~G. Germain, S.~Khurana, and J.~{Le Roux}, ``{SMITIN}: Self-monitored inference-time {IN}tervention for generative music transformers,'' \emph{IEEE Open Journal of Signal Processing}, vol.~6, pp. 266--275, 2025.

\bibitem{facchiano2025activationpatchinginterpretablesteering}
S.~Facchiano, G.~Strano, D.~Crisostomi, I.~Tallini, T.~Mencattini, F.~Galasso, and E.~Rodol{\`a}, ``Activation patching for interpretable steering in music generation,'' \emph{arXiv preprint arXiv:2504.04479}, 2025.

\bibitem{singh2025discovering}
N.~Singh, M.~Cherep, and P.~Maes, ``Discovering and steering interpretable concepts in large generative music models,'' in \emph{Proc. of the Int. Conf. on Learning Representations (ICLR)}, Rio de Janeiro, Brazil, 2026.

\bibitem{zhaosteering2026}
D.~Zhao, D.~Beaglehole, T.~Berg-Kirkpatrick, J.~McAuley, and Z.~Novack, ``Steering autoregressive music generation with recursive feature machines,'' in \emph{Proc. of the Int. Conf. on Learning Representations (ICLR)}, Rio de Janeiro, Brazil, 2026.

\bibitem{jiang2025composer}
X.~Jiang and X.~Xu, ``Composer vector: Style-steering symbolic music generation in a latent space,'' in \emph{NeurIPS 2025 Workshop on AI for Music: Where Creativity Meets Computation}, San Diego, CA, USA, 2025.

\bibitem{staniszewski2026tada}
{\L}.~Staniszewski, K.~Zaleska, M.~Modrzejewski, and K.~Deja, ``{TADA}! {T}uning audio diffusion models through activation steering,'' in \emph{ICLR 2026 Workshop on Representational Alignment (Re-Align)}, Rio de Janeiro, Brazil, 2026.

\bibitem{music21}
M.~S. Cuthbert and C.~Ariza, ``{Music21}: A toolkit for computer-aided musicology and symbolic music data,'' in \emph{Proc. of the 11th Int. Society for Music Information Retrieval Conf. (ISMIR)}, Utrecht, Netherlands, 2010, pp. 637--642.

\bibitem{vig2020investigating}
J.~Vig, S.~Gehrmann, Y.~Belinkov, S.~Qian, D.~Nevo, Y.~Singer, and S.~Shieber, ``Investigating gender bias in language models using causal mediation analysis,'' in \emph{Advances in Neural Information Processing Systems (NeurIPS)}, vol.~33, 2020, pp. 12\,388--12\,401.

\bibitem{geiger2021causal}
A.~Geiger, H.~Lu, T.~Icard, and C.~Potts, ``Causal abstractions of neural networks,'' in \emph{Advances in Neural Information Processing Systems (NeurIPS)}, vol.~34, 2021, pp. 9574--9586.

\bibitem{meng2022locating}
K.~Meng, D.~Bau, A.~Andonian, and Y.~Belinkov, ``Locating and editing factual associations in {GPT},'' in \emph{Advances in Neural Information Processing Systems (NeurIPS)}, vol.~35, 2022, pp. 17\,359--17\,372.

\bibitem{wang2023interpretability}
K.~R. Wang, A.~Variengien, A.~Conmy, B.~Shlegeris, and J.~Steinhardt, ``Interpretability in the wild: A circuit for indirect object identification in {GPT}-2 small,'' in \emph{Proc. of the Int. Conf. on Learning Representations (ICLR)}, Kigali, Rwanda, 2023.

\bibitem{zhang2024towards}
F.~Zhang and N.~Nanda, ``Towards best practices of activation patching in language models: Metrics and methods,'' in \emph{Proc. of the Int. Conf. on Learning Representations (ICLR)}, Vienna, Austria, 2024.

\bibitem{geiger2022inducing}
A.~Geiger, Z.~Wu, H.~Lu, J.~Rozner, E.~Kreiss, T.~Icard, N.~D. Goodman, and C.~Potts, ``Inducing causal structure for interpretable neural networks,'' in \emph{Proc. of the 39th Int. Conf. on Machine Learning (ICML)}, vol. 162, Baltimore, MD, USA, 2022, pp. 7324--7338.

\bibitem{hewitt-manning-2019-structural}
J.~Hewitt and C.~D. Manning, ``A structural probe for finding syntax in word representations,'' in \emph{Proc. of the 2019 Conf. of the North American Chapter of the Association for Computational Linguistics: Human Language Technologies (NAACL-HLT)}, Minneapolis, MN, USA, 2019, pp. 4129--4138.

\bibitem{turner2024steeringlanguagemodelsactivation}
A.~M. Turner, L.~Thiergart, G.~Leech, D.~Udell, J.~J. Vazquez, U.~Mini, and M.~MacDiarmid, ``Steering language models with activation engineering,'' \emph{arXiv preprint arXiv:2308.10248}, 2024.

\bibitem{rimsky-etal-2024-steering}
N.~Rimsky, N.~Gabrieli, J.~Schulz, M.~Tong, E.~Hubinger, and A.~Turner, ``Steering {Llama} 2 via contrastive activation addition,'' in \emph{Proc. of the 62nd Annual Meeting of the Association for Computational Linguistics (Volume 1: Long Papers)}, Bangkok, Thailand, 2024, pp. 15\,504--15\,522.

\bibitem{retkowski2024frechet}
J.~Retkowski, J.~St{\k{e}}pniak, and M.~Modrzejewski, ``Fr{\'e}chet music distance: A metric for generative symbolic music evaluation,'' \emph{arXiv preprint arXiv:2412.07948}, 2024.

\end{thebibliography}

\end{document}